\documentclass[aps,epsf,twocolumn,showpacs]{revtex4}
\usepackage{amsmath}
\usepackage{amssymb}
\usepackage{epsfig}
\usepackage{graphicx}
\usepackage{color}
\usepackage[normalem]{ulem}

\begin{document}

\title{Parallel multicanonical study of the three-dimensional Blume-Capel model}

\author{Johannes Zierenberg$^1$}

\author{Nikolaos G. Fytas$^2$}

\author{Wolfhard Janke$^1$}

\affiliation{
$^1$Institut f\"{u}r Theoretische Physik, Universit\"{a}t Leipzig, Postfach 100
920, D-04009 Leipzig, Germany\\
$^2$Applied Mathematics Research Centre, Coventry University, Coventry, CV1 5FB,
United Kingdom
}

\date{\today}

\begin{abstract}
We study the thermodynamic properties of the three-dimensional
Blume-Capel model on the simple cubic lattice by means of computer
simulations. In particular, we implement a parallelized variant of the
multicanonical approach and perform simulations by keeping a constant
temperature and crossing the phase boundary along the crystal-field
axis. We obtain numerical data for several temperatures in both the
first- and second-order regime of the model. Finite-size scaling
analyses provide us with transition points and the dimensional scaling
behavior in the numerically demanding first-order regime, as well as a
clear verification of the expected Ising universality in the
respective second-order regime. Finally, we discuss the scaling
behavior in the vicinity of the tricritical point.
\end{abstract}

\pacs{75.10.Nr, 05.50.+q, 64.60.Cn, 75.10.Hk} \maketitle
\newcommand{\Emuca}{E_{\Delta}}

\section{Introduction}
\label{sec:Intro}

The Blume-Capel (BC) model consisting of a spin-one Ising Hamiltonian with a
single-ion uniaxial crystal field anisotropy~\cite{blume66,capel66} is one of
the most studied models in the communities of Statistical Mechanics and
Condensed Matter Physics. This is not only because of the relative simplicity
with which approximate calculations for this model can be carried out and
tested, as well as the fundamental theoretical interest arising from the
richness of its phase diagram, but also because versions and extensions of the
model can be applied for the description of many different physical systems,
some of them being multi-component fluids, ternary alloys, and \mbox{$^{3}$He
-- $^{4}$He} mixtures~\cite{lawrie}. Recent applications of the BC model
include analyses of ferrimagnets, as discussed in a thorough contribution by
Selke and Oitmaa~\cite{selke-10}.

The BC model is described by the Hamiltonian
\begin{equation}
\label{eq:ham}
\mathcal{H}=-J\sum_{\langle ij\rangle}\sigma_{i}\sigma_{j}+\Delta\sum_{i}\sigma_{i}^{2}=E_{
J}+\Delta \Emuca,
\end{equation}
where the spin variables $\sigma_{i}$ take on the values $\{-1, 0,+1\}$,
$\langle ij\rangle$ indicates summation over all nearest-neighbor pairs of
sites, and $J>0$ is the ferromagnetic exchange interaction (here we set $J=1$
and $k_{\rm B}=1$ to fix the temperature scale). The parameter $\Delta$ is
known as the crystal-field coupling that controls the density of vacancies
($\sigma_{i}=0$). For $\Delta\rightarrow -\infty$ vacancies are suppressed and
the model maps onto the Ising model. We always employ periodic boundary
conditions. Note here that the second formulation of the
Hamiltonian~(\ref{eq:ham}), via the definitions of $E_{J}$ and $E_{\Delta}$,
will allow us to define the necessary observables for the application of our
finite-size scaling (FSS) scheme that will be discussed in detail below.

\begin{figure}[htbp]
\includegraphics*[width=8 cm]{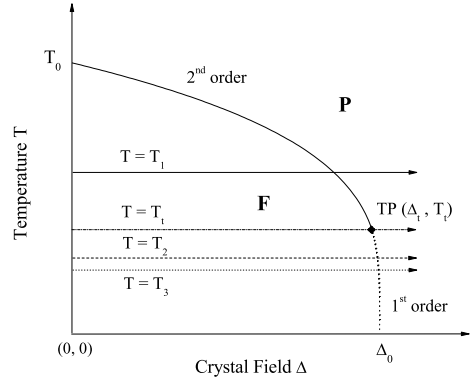}
\caption{\label{fig:phd} General sketch of the phase diagram of
  the 3D Blume-Capel model in the temperature -- crystal field plane.  The
  phase boundary separates the ferromagnetic (\textbf{F}) from the paramagnetic
  (\textbf{P}) phase, in which the solid line indicates continuous and the
  dotted line first-order phase transitions, respectively. The two lines merge
  at the tricritical point (TP), as highlighted by the black rhombus. The
  limiting cases of $T=0$ and $\Delta=0$ are marked on the relevant axis with
  $\Delta_{0}=3$ and $T_{0} = 3.195(1)$~\cite{fytaspanos}, respectively. The horizontal
  arrows illustrate the direction of crossing the phase boundary at fixed
  temperatures, studied in this work, in both the second-order ($T_{1}=2.0$)
  and first-order ($T_{2}=1.0$ and $T_{3}=0.9$) regimes, as well as in the
  vicinity of the tricritical point ($T_{\rm t}=1.4182$)~\cite{deserno97}.
}
\end{figure}

As is well known, the pure and disordered versions of the model of
Eq.~(\ref{eq:ham}) have been analyzed, besides the original
mean-field theory~\cite{blume66,capel66}, by a variety of
approximations and numerical approaches, in both two (2D) and
three dimensions (3D). These include the real-space
renormalization group~\cite{branco}, Monte Carlo
renormalization-group calculations~\cite{landau72},
$\epsilon$-expansion renormalization groups~\cite{stephen73},
high- and low-temperature series calculations~\cite{fox73}, a
phenomenological FSS analysis using a strip
geometry~\cite{nightingale82,beale86}, and of course Monte Carlo
simulations~\cite{jain80,landau81,care93,deserno97,Blote95,Blote04,silva06,malakis09,malakis12,theodorakis12,fytas,fytaspanos}.

The phase diagram of the pure model consists of a segment of
continuous Ising-like transitions at high temperatures and low
values of the crystal field. This ends at a tricritical point
($\Delta_{\rm t},T_{\rm t}$) where it is joined with a second
segment of first-order transitions ending for $T=0$ at $\Delta_{0} =
zJ/2$, with $z$ denoting the coordination number of the considered
lattice. In the present case of a simple cubic lattice, where $z=6$,
it follows that $\Delta_{0} = 3$. A general sketch of the phase
diagram is given in Fig.~\ref{fig:phd} and is outlined in the relevant
caption. The location of the tricritical point, marked by the black
rhombus in Fig.~\ref{fig:phd}, has been estimated by
Deserno~\cite{deserno97} using a microcanonical Monte Carlo approach
to be $(\Delta_{\rm t},T_{\rm t}) = (2.84479(30),1.4182(55))$.

The scope of the present paper is to present a complementary study
of the 3D BC model embedded in the simple cubic lattice. Our
simulations follow a sophisticated numerical scheme, outlined in
the following Sec.~\ref{sec:MUCA}, using as a platform the
multicanonical approach. This is especially suitable for the
study of systems that undergo a first-order phase transition,
where it is well-known that numerical simulation is a hard task
to perform. One interesting aspect of the present work is that we
cross, in our simulations, the phase boundary of the system along
the crystal-field axis, keeping the temperature fixed. Thus we
obtain relevant thermodynamic observables as a function of the
crystal field $\Delta$ and we perform a FSS analysis on a
different basis. This analysis is presented in Sec.~\ref{sec:FSS},
in both the first- and second-order regimes of the model, where
the second-order regime is used as a test case of our scheme and
more attention is paid to the first-order regime of the phase
diagram, which is computationally much more challenging. In
particular, we consider three different temperatures, one in the
second-order regime, $T_{1}=2.0$, and two of them in the first-order
regime, $T_{2}=1.0$ and $T_{3}=0.9$. Moreover, we discuss the scaling
properties in the vicinity of the proposed tricritical point, by
performing additional simulations and relevant analysis at the
temperature $T_{\rm t}=1.4182$ suggested in Ref.~\cite{deserno97}.
This contribution is ended in Sec.~\ref{sec:Conclusions}, where a
brief summary of our conclusions is given together with an outlook for
future work.

\section{Numerical method and scaling observables}
\label{sec:MUCA}

We apply a multicanonical method~\cite{berg1992, janke1998} with
the slight modification to yield a flat histogram not in the total
energy $E$, but rather in $E_{\Delta}$. The multicanonical method
allows one to increase the probability to sample otherwise suppressed
states and, with it, overcome emerging barriers. Hence, it is an
optimal tool to study first-order phase transitions. The canonical
expectation value weights all observables of the phase space with
the Boltzmann weight, which leads to the general form
\begin{equation}
    \langle O\rangle = \frac{1}{Z}\sum_{x}O(x)~e^{-\beta E(x)},
  \label{eq:ExpValueCan}
\end{equation}
where $Z=\sum_{x}~e^{-\beta E(x)}$ is the partition sum, $\beta=1/T$ is
the inverse temperature, and $x$ stands short for the spin configurations.
For the usual multicanonical method, the Boltzmann weight in the canonical
probability distribution $\exp\{-\beta E(x)\}$ is replaced by a weight
function $W(E(x))$, that is iteratively modified to yield a flat energy histogram.
At this point, we can rewrite $E=E_{J}+\Delta \Emuca$, separate the probability
distribution, and replace the Boltzmann weight depending on $\Emuca$:
\begin{equation}
  e^{-\beta E_{J}}~e^{-\beta\Delta \Emuca}
\rightarrow e^{-\beta E_{J}}~W\left(\Emuca\right).
\end{equation}

\begin{figure*}[htbp]
\includegraphics[width=8 cm]{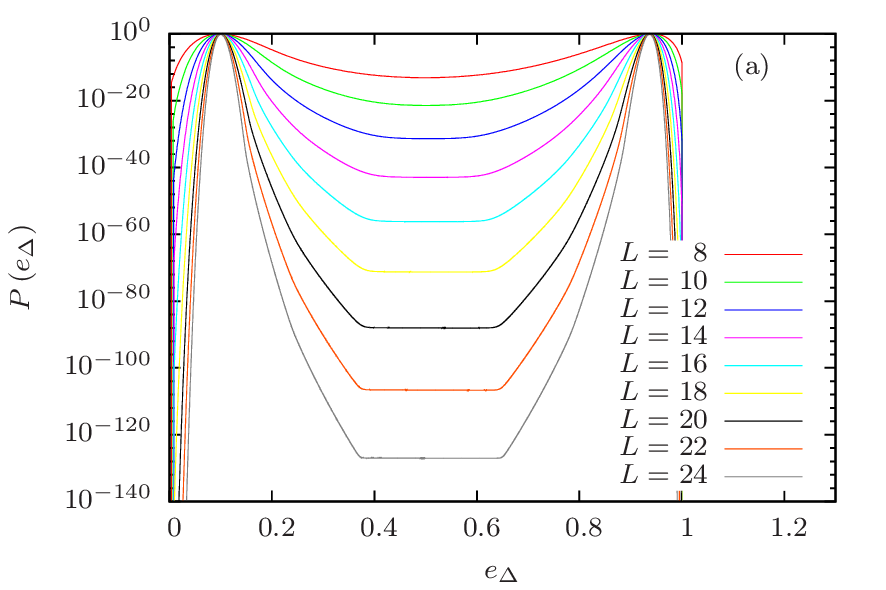}
\includegraphics[width=8 cm]{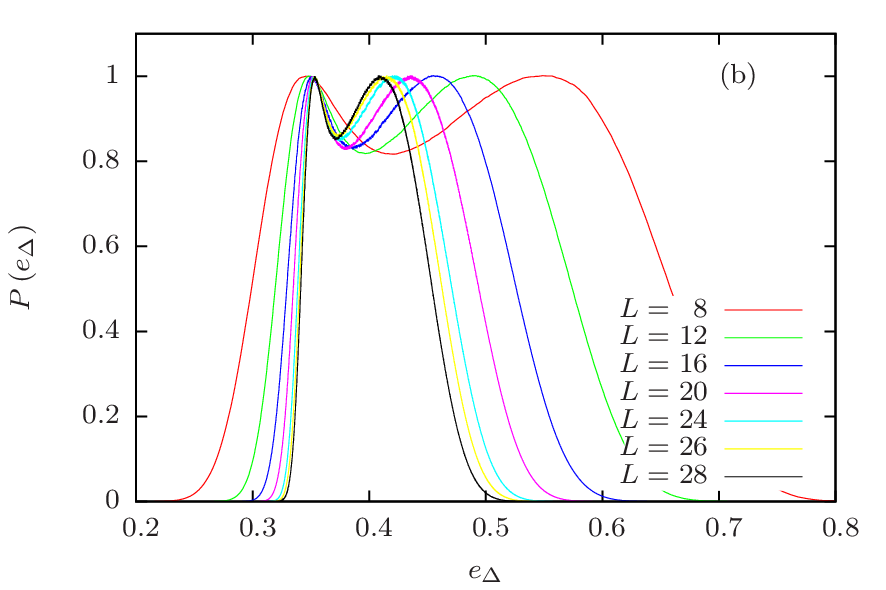}
\caption{\label{fig:barrier}
  (color online) Illustration of the reweighted probability distribution
  at the transition field $\Delta_{\rm
eqh}$ with respect to $e_{\Delta} = E_{\Delta}/V$ at (a) $T=0.9$,
that is well inside the first-order regime of the system (note the
strongly increasing barrier with the system size), exhibiting a major
suppression of intermediate states which is common for first-order
transitions, and (b) at $T=T_{\rm t}=1.4182$ which is in the
vicinity of the tricritical point.}
\end{figure*}

Considering a fixed inverse temperature $\beta$, one is then able to
iteratively adapt $W\left(E_{\Delta}\right)$ in order to yield a flat
histogram in $E_{\Delta}$. This is in fact quite similar to
multimagnetic simulations and also suited for the application of a
parallel implementation of the multicanonical
method~\cite{zierenberg2013}. We made use of this parallelization with
up to $36$ cores, which speeds up the iteration process and provides
$36$ independent production runs. The canonical expectation values at
a certain point ($\beta$, $\Delta$) may then be estimated with
standard histogram and time-series reweighting
techniques~\cite{reweighting}. Since the multicanonical simulation is
still an importance sampling Markov chain, one only needs to consider
the multicanonical variable $E_{\Delta}$, illustrated for the case of
time-series reweighting of a given observable $O$:
\begin{equation}
  \langle O \rangle_{\beta,\Delta}
  = \frac{\sum_{x} O(x) e^{-\beta\Delta \Emuca(x)}W^{-1}\left(\Emuca(x)\right)}
         {\sum_{x}      e^{-\beta\Delta \Emuca(x)}W^{-1}\left(\Emuca(x)\right)}.
\end{equation}
Here, $\langle \cdots \rangle$ clearly refers to the estimator of
the expectation value and we will drop the subscripts in the
following. Figure~\ref{fig:barrier} shows an example of the
reweighted probability distributions of $e_{\Delta}=\Emuca/V$,
where $V=L^{3}$ and $L$ denotes the linear system size, at the
transition field $\Delta_{\rm eqh}$, i.e. where the distribution
shows two peaks of equal height. Well inside the first-order regime
the system shows a barrier increasing with the system size,
characteristic of the nature of the transition, which reaches at
$T_3=0.9$ already for a system size of $V=24^3$ spins the order of
$10^{-130}$ , see Fig.~\ref{fig:barrier}~(a).  On the other
hand, at the proposed tricritical point (see panel (b) with numerical
data obtained at $T=T_{\rm t}=1.4182$~\cite{deserno97}) no definite
judgement can be made. We observe that the distributions still show a
double-peaked structure, yet with a much smaller barrier, which does
not diverge with increasing system size. A more illuminating
discussion of this special temperature is given below at the end of
Sec.~\ref{sec:FSS}.

It is worth noting here that the $\Emuca$ are always integer in
the range $[0,V]$, which is a major advantage for the application
of the present method. This would not be the case for the usual
multicanonical~\cite{berg1992,janke1998} and Wang-Landau~\cite{WL}
methods at fixed $\Delta$ based on the total energy $E$, because $E
\in \mathbb{R}$ for non-integer values of $\Delta$. Additionally, our
previous experience with the application of the Wang-Landau method to
the BC model in the low-temperature regime for a fixed value of
$\Delta$ suggested that one needs to be extremely careful in the
details of implementation. For instance, the need to simulate large
enough system sizes leads to the inevitable application of a
multi-range approach of the Wang-Landau method, that is splitting the
energy range in many subintervals~\cite{malakis09}. This approach,
although appearing to be much faster than the straightforward
one-range implementation, may give rise to several problems with
respect to the breaking of ergodicity of the
process~\cite{troyer03,dayal04,belardinelli07} and possible
distortions (systematic errors) induced on the density of
states~\cite{fytasme}. On the other hand, the numerical framework
developed and applied in the current paper does not suffer from this
type of inherent problems. On the contrary the parallelized variant of
the multicanonical approach used, combined with the orthogonal scaling
of the phase boundary has proven to be a promising scheme for the
first-order transition regime. This will be clearly shown in the
following section with the accurate estimation of transition points
$\Delta^{\ast}$ even for temperatures $T < 1$, which is a, commonly
accepted, harsh numerical task. Still, for the second-order regime
both the current approach and any other type of generalized ensemble
sampling method would give comparable results within the statistical
errors as we have already verified by our preliminary numerical tests.
In fact, the modification to a flat-histogram method in a sub-energy
is not restricted to the present model or a spin system in general and
has been applied in a similar way also to a polymer system in
disorder~\cite{Schoebl2011}. Moreover, the formulation for other
generalized ensemble methods is straight forward.

In order to obtain transition points for the FSS analysis, one usually
considers the peak of the specific heat $C$, magnetic susceptibility
$\chi$, or any other suitable temperature derivative of an order
parameter~\cite{reweighting}. In principle, the magnetic properties of
the system show a more reliable behavior when one is interested in
obtaining accurate estimates of critical points and it is a common
practise along these lines to firstly estimate the magnetic exponents
$\beta$, $\gamma$, and $\nu$, and then via the hyperscaling relation the
exponent $\alpha$ of the specific heat. However, also other scaling
approaches based on a different philosophy have been successfully used
in the literature, depending always on the direction of intersecting
the phase boundary of the model under study. For instance, it has been
shown that for the 3D random-field Ising model at zero temperature,
the field derivative of the bond energy $E_{J}$ defines a
specific-heat-like quantity from which one may produce accurate
estimates of the critical exponent ratio $\alpha/\nu$ and whose shift
behavior defines correct critical points in the field -- temperature
plane~\cite{hartmann}.

For the present study, where we are crossing the phase diagram at
fixed temperature along the crystal-field axis, we may as well
consider instead of the standard definitions, the field derivative
of the form $\partial/\partial\Delta$. The derivative of the
expectation value \eqref{eq:ExpValueCan} yields
\begin{align}
  \frac{\partial\langle O\rangle}{\partial\Delta}
  & = -\beta [\langle O\Emuca\rangle - \langle
O\rangle\langle\Emuca\rangle] + \left\langle
\frac{\partial}{\partial\Delta}O\right\rangle,
\label{eq:FieldDerivative}
\end{align}
which is similar to any specific-heat-like quantity because in
general the observable is independent on the variable and the last
term drops out. This is true for either $E_i$, where $i=J$ or
$\Delta$, however, the total energy $E=E_{J}+\Delta \Emuca$ is no
longer independent on the field which leads to
\begin{align}
  \frac{\partial\langle E\rangle}{\partial\Delta}
  & = -\beta \big[\langle E\Emuca\rangle - \langle
  E\rangle\langle\Emuca\rangle\big]
      + \langle\Emuca\rangle \nonumber\\
  & = \frac{\partial \langle E_J\rangle}{\partial\Delta}
      +\Delta\frac{\partial \langle \Emuca\rangle}{\partial\Delta}
      + \langle\Emuca\rangle.
\end{align}
However, the last line suggests -- expecting a critical or
diverging behavior -- that we may limit our consideration to either
of the energy contributions. In fact, we consider here only the
field derivative of the spin-spin interaction term
\begin{equation}
\label{eq:sph}
  C(\Delta) = \frac{\partial \langle E_J\rangle}{\partial\Delta}
            =  - \beta \big[\langle E_J\Emuca\rangle - \langle
E_J\rangle\langle\Emuca\rangle\big].
\end{equation}
Similar considerations may apply also for other suitably defined
thermodynamic functions that could provide us, for instance, with
estimates of magnetic exponents mentioned above. Yet, this task goes
beyond the scope of the present work where we focus on the first-order
transition regime of the BC model and target only at a qualitative
comparison to the expected Ising criticality in the second-order
transition regime. In fact, this comparison becomes even more clear
via the use of the straightforwardly defined specific-heat-like
quantity~(\ref{eq:sph}).

\begin{figure}[tbp]
\includegraphics*[width=8 cm]{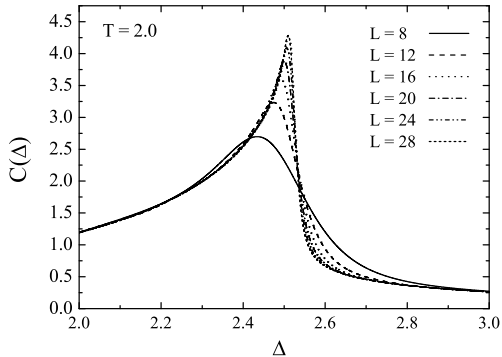}
\caption{\label{fig:sphT2} Specific-heat curves as a function of
the crystal field $\Delta$ for $T = 2.0$ and several system sizes
in the second-order regime. Smooth curves typical of continuous
transitions with a clear shift behavior are observed.}
\end{figure}

Relevant plots of $C(\Delta)$ can be found in the next section in
Figs.~\ref{fig:sphT2} and \ref{fig:sphT1} and will be discussed
there. However, it is obvious from these illustrations that there
exists clearly a maximum of $C(\Delta)$ that moreover shows a
shift behavior as well. Let us define now $\Delta^{\ast}_{L}$ as
the crystal field value at which $C(\Delta)$ attains its maximum,
and as we shall see below this defines a suitable pseudocritical,
or pseudotransition, parameter that carries in itself the approach
to the thermodynamic limit in the second- and first-order regime
of the model, respectively. Similarly, we denote by $C^{\ast}_{L}$
the value of the specific heat at this pseudocritical point
$C(\Delta^{\ast}_{L})$.  The value of $\Delta^{\ast}_{L}$ is
numerically determined by calculating the second derivative of
$\langle E_J\rangle$ with respect to $\Delta$, analogous to the
above calculations, and finding the zero crossing. For this, we
apply a bisection algorithm with time-series reweighting for the
full data set as well as $w$ subsets, each excluding $1/w$
measurements, for the jackknife error
calculation~\cite{efron1982}. The jackknife method is similarly
then used for $C^{\ast}_{L}$.

To sum up, using the above scheme we have performed simulations
for the three temperatures shown in Fig.~\ref{fig:phd} and
outlined in the introduction, as well as for the tricritical
temperature $T_{\rm t}=1.4182$~\cite{deserno97}. In all cases, we
considered various linear sizes within the range $L=8-28$. In
principle, our numerical approach could as well simulate even
larger system sizes, especially in the second-order regime. Still
we have found it useful to optimize our code following the needs
of a careful inspection of the first-order regime, for which
linear sizes of the order of $L=28$ are already quite large,
taking into account that we are well into the low-temperature part
of the phase diagram of the model.

\begin{figure}[htbp]
\includegraphics*[width=8 cm]{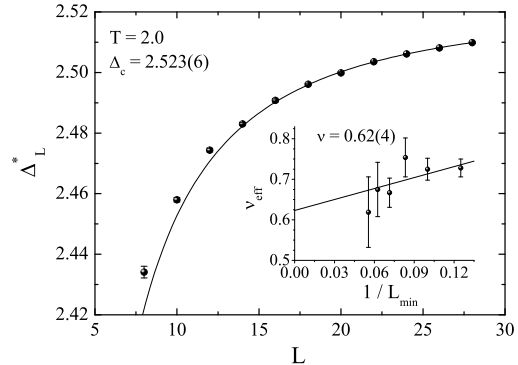}
\caption{\label{fig:scalingT2}
  Shift behavior~(\ref{eq:shift}) of the pseudocritical fields
  $\Delta_{L}^{\ast}$ at $T=2.0$ obtained from the peak location of the specific-heat curves
  shown in Fig.~\ref{fig:sphT2}. The solid line is a fit for linear sizes $L\ge L_{\rm min}=18$,
  giving a stable, under $L_{\rm min}$
  changes, critical field value $\Delta_{\rm c}=2.523(6)$. The inset illustrates
  the extrapolation of the effective exponent $\nu$, obtained by varying the
  cutoff $L_{\rm min}$ of the fits, to the thermodynamic limit.
}
\end{figure}

\section{Finite-size scaling analysis}
\label{sec:FSS}

In this section, we present the main results of our contribution based
on a FSS analysis of the numerical data obtained with the method
outlined above. As a first step we shall consider the scaling in the
second-order regime of the model ($T > T_{\rm t}$) and in particular
at the temperature $T_{1}=2.0$. Let us point out here that when it
comes to the second-order transition regime of the BC model, the
modified multicanonical method as implemented here is by no means the
method of choice if one wants to obtain high-accuracy estimates of
critical exponents (or critical points) and cannot compete against
other cluster-update methods~\cite{hasen} or more involved generalized
ensemble schemes~\cite{BergMulti,Bittner2011} especially tailored to
this situation. However, a qualitative study at this regime allows a
direct comparison to the extensive and precise literature of the
simple 3D Ising model, thus serving as a clear cut test of the
proposed scheme. Additionally it justifies the results and conclusions
drawn from our study at the first-order regime, presented later in
this section for two temperature values ($T_{2}=1.0$ and $T_{3}=0.9$
in Fig.~\ref{fig:phd}), which is not so easy to control given the huge
energy barriers illustrated in Fig.~\ref{fig:barrier}(a). Finally, in
the last part of this section, we will discuss the scaling behavior of
our observables at an estimate of the tricritical
point~\cite{deserno97}.

\begin{figure}[tbp]
\includegraphics*[width=8 cm]{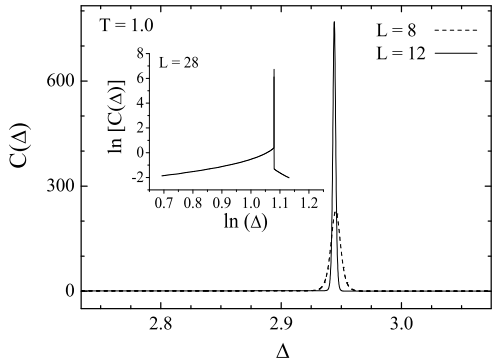}
\caption{\label{fig:sphT1}
  Specific-heat curves as a function of the crystal field
  $\Delta$ at $T=1.0$ and two system sizes $L=8$ (dashed line) and $12$ (solid line). The inset illustrates
  a typical first-order-like specific-heat curve for a system with linear size
  $L=28$ on a double logarithmic scale.
}
\end{figure}

Figure~\ref{fig:sphT2} displays the specific-heat-like curves defined
in Eq.~(\ref{eq:sph}) as a function of the crystal field $\Delta$ for
$T=2.0$. Several system sizes up to $L=28$ are shown which exhibit a
clear shift behavior. This is further quantified in
Fig.~\ref{fig:scalingT2}, which presents the FSS behavior of the
pseudocritical fields $\Delta_{L}^{\ast}$ estimated as the locations
where the specific heat attains its maximum. As usual, for
second-order phase transitions a scaling behavior of the form
\begin{equation}
  \label{eq:shift}
  \Delta_{L}^{\ast} = \Delta_{\rm c} + b L^{-1/\nu}
\end{equation}
is used in order to describe the approach to the thermodynamic
limit.
It appears that this method of extracting pseudo-critical points
from the maxima of some properly defined thermodynamic quantity is
capable of producing accurate estimates for both the critical
crystal field $\Delta_{\rm c}$ and also the correlation-length
exponent $\nu$, assuming that its behavior follows the observed
shift behavior of our pseudocritical fields $\Delta_{L}^{\ast}$.
It is well known from the general scaling theory that, even for
simple models, the equality between the correlation-length
exponent and the shift exponent is not a necessary consequence of
scaling~\cite{barber83}. Of course, it is a general practice to
assume that the correlation-length behavior can be deduced by
the shift of appropriate thermodynamic functions.

In fact, the solid line in the main panel of
Fig.~\ref{fig:scalingT2} represents a fitting of the
form~(\ref{eq:shift}), using as a lower cutoff the linear size
$L_{\rm min} = 18$. We have performed this type of analysis for
several values of $L_{\rm min}$ within the range $8 - 18$ and
keeping of course the upper system size fixed at $L_{\rm max} =
28$. For each of these fits we have estimated an effective value
of the correlation-length's exponent which is plotted in the inset
of Fig.~\ref{fig:scalingT2} as a function of the inverse lower
cutoff, i.e., the parameter $1/L_{\rm min}$. A linear
extrapolation to the infinite-limit size provides an estimate of
$\nu=0.62(4)$, which within error bars is compatible with the
Ising universality exponent $\nu=0.6304(13)$~\cite{guida}, as
expected. Regarding the value of the critical field we obtain the
estimate $\Delta_{\rm c}(T_1=2.0) = 2.523(6)$, that remained quite stable
under the switching of the lower cutoff during the fitting
procedure. Thus, up to this point we have verified through a
rather different, ``orthogonal'' route the expected Ising
universality in the second-order phase transition regime of the 3D
BC model.

\begin{figure}[htbp]
\includegraphics*[width=8 cm]{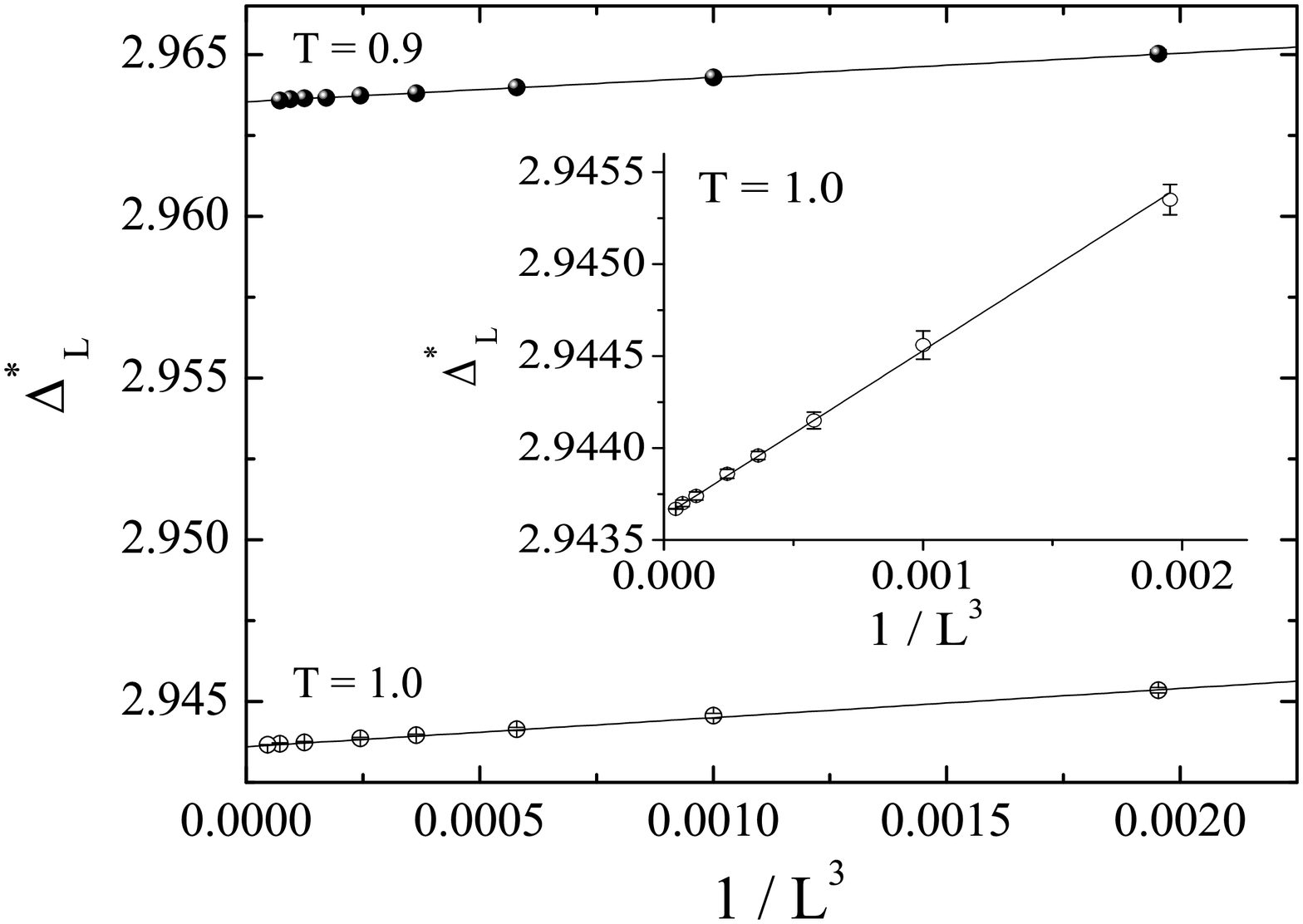}
\caption{\label{fig:scalingT1}
  Shift behavior~(\ref{eq:shift_new}) of the pseudocritical fields
  $\Delta_{L}^{\ast}$ obtained from the peak location of the curves (a sample of
  which is shown in Fig.~\ref{fig:sphT1}) for both temperatures $T=1.0$ and
  $T=0.9$ in the first-order regime of the phase diagram. The obtained
  transition fields $\Delta^{\ast}$ are given in the main text. The inset is a
  mere enlargement of the $\sim L^{-d}$ approach of the pseudotransition points
  $\Delta_{L}^{\ast}$ to $L\rightarrow \infty$ for the temperature $T=1.0$.
}
\end{figure}
We now move on to the main objective of this work, the discussion of
the characteristics of the transition in the first-order regime and
its dimensional scaling behavior. Let us point out here before
discussing our findings that crossing the boundary at the first-order
transition regime at a fixed temperature is an orthogonal approach to
the fixed-field ansatz.  One advantage in the case of the BC model is
a broad temperature range with a first-order transition in comparison
to a small $\Delta$-range. 

As we already discussed above, we have obtained numerical data at
two temperatures in the first-order regime of the model,
$T_{2}=1.0$ and $T_{3}=0.9$. A nice illustration of the
first-order character of the transition at these temperatures is
shown in Fig.~\ref{fig:sphT1}, where we plot the specific-heat
curves obtained from Eq.~(\ref{eq:sph}) for $T=1.0$ and two system
sizes, $L=8$ and $L=12$ (and $L=28$ in the inset at a double
logarithmic scale). Clearly, a sharp peak is observed which
becomes much more pronounced with increasing system size.

Following a similar analysis as above, we study now the scaling of the
pseudocritical fields $\Delta_{L}^{\ast}$ obtained from the sharp
specific-heat peaks. In this case we would expect a scaling of the
form
\begin{equation}
  \label{eq:shift_new}
  \Delta_{L}^{\ast} = \Delta^{\ast} + b L^{-d},
\end{equation}
where $d=3$ the dimensionality of the lattice and $\Delta^{\ast}$
the transition field. The above shift behavior of the
pseudocritical fields $\Delta_{L}^{\ast}$ for both temperatures
$T=1.0$ and $T=0.9$ in the first-order regime of the phase diagram
is shown in Fig.~\ref{fig:scalingT1} as a function of the inverse
of the volume of the system. The solid lines are linear
extrapolations to the infinite-limit size (for a clearer
illustration of the linear behavior see the corresponding inset).
The obtained transition fields $\Delta^{\ast}$ are $2.944(5)$ and
$2.964(6)$ for $T=1.0$ and $0.9$, respectively.

\begin{figure}[t]
\includegraphics*[width=8cm]{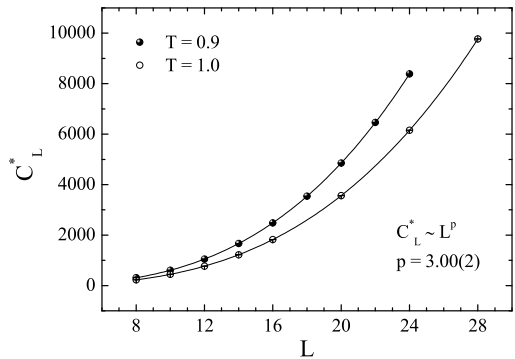}
\caption{\label{fig:scalingsphT1}
  Simultaneous fitting of the specific-heat maxima at both temperatures in the
  first-order regime. The expected $\sim L^{d}$ scaling behavior is obtained as
  can be clearly seen.
}
\end{figure}

As another important aspect of the first-order regime in the phase
diagram of the model, we study the scaling of the specific-heat maxima
in Fig.~\ref{fig:scalingsphT1}. In particular, we plot the FSS
behavior of the peaks for both temperatures considered in this regime,
where the two solid lines show a simultaneous fitting attempt of the
form $C_{L}^{\ast} \sim L^{p}$, simultaneous meaning that the two data
sets share the same exponent during the fitting procedure. Of course,
in a standard first-order phase transition, the exponent $p$ is
expected to be equal to the dimensionality $d$ of the system, that is
$3$ in our case. The result for the exponent $p$ of a simultaneous fit
to the data for both temperatures with $\chi^{2}/{\rm dof}\approx 0.8$
is $p = 3.00(2)$, which is in excellent agreement with the theoretical
expectation $p=d=3$.

Further to the above successful study of criticality in the
second-order regime of the model, it is now clear that the numerical
method and scaling approach implemented in the present paper is able
to capture as well the first-order characteristics of the transition
within a good accuracy. This latter fact is of particular importance
as we are dealing with the low-temperature first-order regime of the
BC model, where it is common knowledge that most numerical methods
fail to produce reliable estimates of transition points and
criticality. Thus, the current method could be easily stretched to
produce an accurate approximation of the phase boundary line for
values of the crystal field $\Delta$ within the regime $[\Delta_{\rm
t}, 3]$.

\begin{figure}[t]
  \centering
  \includegraphics[]{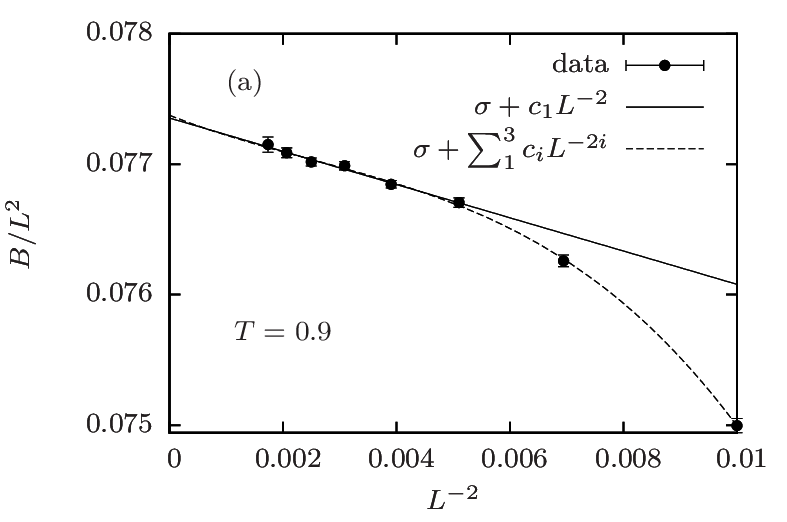}\\
  \includegraphics[]{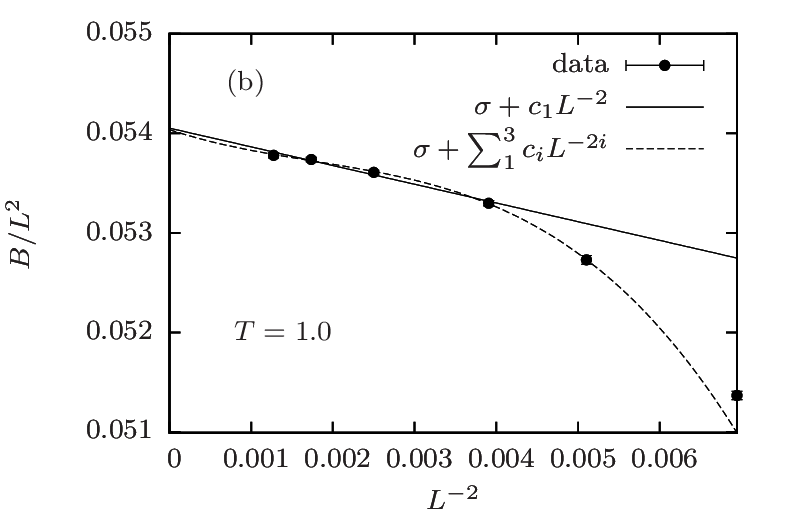}\\
  \includegraphics[]{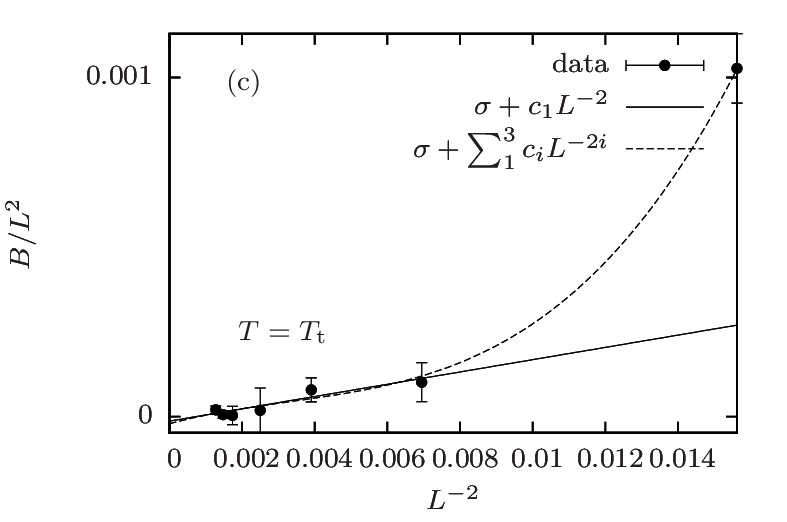}
  \caption{\label{fig:barrierScaling}
      Scaling of the barrier height $B$ for all considered temperatures $T\leq
      T_{\rm t}$. This barrier may be associated with the transition from a
      spin-$0$ dominated to a spin-$\pm1$ dominated regime (see
      Fig.~\ref{fig:barrier}). Then $B/L^2$ plays the role of an interface
      tension for spin-$0$ strips.
  }
\end{figure}

The multicanonical method allows us to directly estimate the barrier associated
with the suppression of states during the first-order phase transition, as shown
in Fig.~\ref{fig:barrier}.  Considering distributions with two peaks of equal
height, i.e., two equally probably states, leads to the formulation of a
free-energy like barrier in the $\Emuca$-space,
\begin{equation}
  B = \frac{1}{2\beta\Delta}\ln\left(\frac{P_{\rm
  max}}{P_{\rm min}}\right)_{\rm eqh},
\end{equation}
where $P_{\rm max}$ and $P_{\rm min}$ are the maximum and the
local minimum of the distribution $P(\Emuca)$, respectively. The
resulting barrier connects a spin-$0$ dominated regime ($\Emuca$
small) and a spin-$\pm1$ dominated regime ($\Emuca$ large). This
shows large similarities to the Ising (lattice gas) model and the
according droplet/strip
transitions~\cite{Nussbaumer2006,Nussbaumer2008}. Thus, the
association with condensation and strip formation of spin-$0$
clusters seems natural and we would expect a scaling behavior in
three dimensions as $B / L^2 = \sigma + c_1L^{-2} +
\mathcal{O}\left(L^{-4}\right)$ possibly with higher-order
corrections~\cite{Bittner2009}. Figure~\ref{fig:barrierScaling}
shows $B/L^2$ as a function of $L^{-2}$ for $T_{3}=0.9$, $T_2=1.0$
and $T_{\rm t}=1.4182$ with fits of the data including the first
correction and up to the third corrections.
While higher-order corrections describe the systematic dependence
of the data better, the $L\rightarrow\infty$ extrapolations are
consistent within error bars for both fits yielding the estimates
$\sigma_3=0.0774(1)$ and $\sigma_2=0.0540(2)$ for $T_3=0.9$  and
$T_2=1.0$, respectively. In the vicinity of the tricritial point,
at $T_{\rm t}=1.4182$, the extrapolation yields $\sigma \approx 0$
indicating as expected that the interface tension vanishes in the
thermodynamic limit.
\begin{figure}[t]
\includegraphics*[width=7 cm]{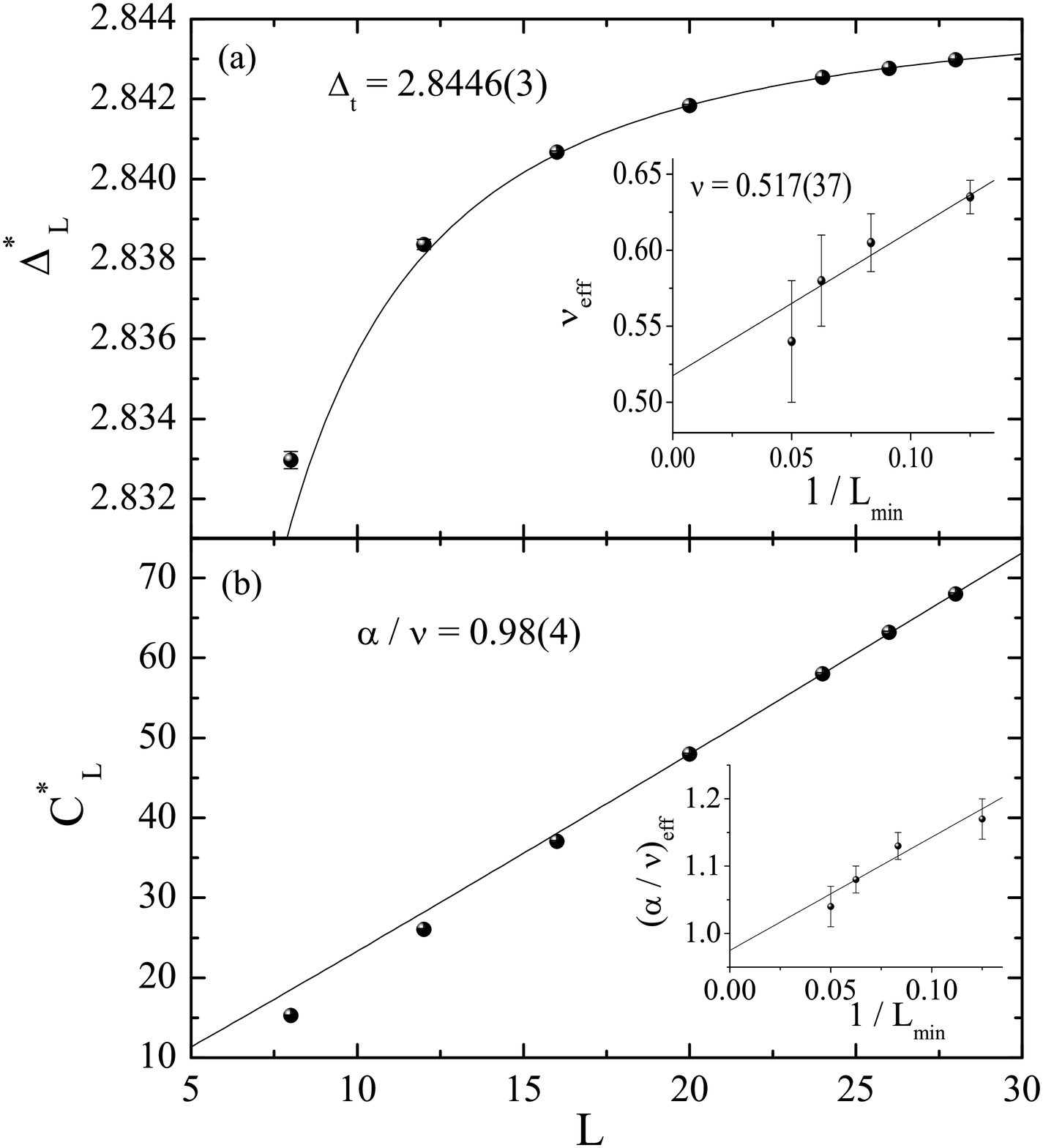}
\caption{\label{fig:tricritical}
  Critical aspects of the 3D Blume-Capel model at the tricritical point proposed in
  Ref.~\cite{deserno97}: $T=T_{\rm t}=1.4182$. (a) Shift  behavior of
  $\Delta_{L}^{\ast}$ obtained from the location of the specific-heat peaks. The
  solid line is a power-law fit of the form~(\ref{eq:shift}) for $L\geq L_{\rm
  min}=20$. The inset illustrates the infinite-volume extrapolation of the
  correlation-length effective exponent by varying the lower cutoff $L_{\rm min}$
  during the fittings. (b) Scaling behavior of the specific-heat peaks again
  for the larger system sizes. The corresponding inset shows an infinite-volume
  extrapolation of the effective exponent ratio $\alpha/\nu$ using the same
  procedure as in (a).
}
\end{figure}

In the last part of this section we discuss some scaling results
in the vicinity of the tricritical point of the 3D BC
model~\cite{lawrie}. We have performed additional simulations by
fixing the temperature at the tricritical estimate $T=T_{\rm
t}=1.4182$, as suggested by Deserno~\cite{deserno97}, crossing
again the phase diagram along the crystal-field axis. The results
and relevant FSS analysis are given in Fig.~\ref{fig:tricritical},
where one can clearly observe the departure from the Ising
second-order universality class to the tricritical one, at least
in terms of the estimated critical exponents. In particular, in
panel (a) of this figure we present the shift behavior of
$\Delta_{L}^{\ast}$ obtained from the location of the
specific-heat peaks at the above defined temperature. The solid
line is a power-law fitting of the form~(\ref{eq:shift}) for
$L\geq L_{\rm min}=20$ and the estimate we obtain for the relevant
(tricritical) crystal-field value is $\Delta_{t}=2.8446(3)$. This
latter value compares very well to the value $2.84479(30)$
proposed by Deserno, using an empirical scaling of the coordinates
of a latent-heat-like quantity of the model. The inset of panel
(a) illustrates correspondingly the infinite-limit size
extrapolation of the correlation-length's effective exponent by
varying the cutoff $L_{\rm min}$ during the fitting procedure, as
also performed in the analysis within the second-order regime of
the model (see Fig.~\ref{fig:scalingT2}).  The obtained value of
$\nu$, that is $\nu = 0.517(37)$, is clearly different to that of
the standard second-order Ising universality class, and within
error bars, compatible to the theoretical expectation of the Ising
tricritical universality value of $\nu =
0.5$~\cite{lawrie,Blote04}. This result indicates that the
estimate of Deserno~\cite{deserno97} for the location of the
tricritical point in the \mbox{temperature -- crystal-field} plane
is indeed quite accurate, and secondly it provides a strong test
in favor of the implemented numerical and scaling scheme of the
present paper. Further to these results, we present in
Fig.~\ref{fig:tricritical}~(b) the scaling behavior of the
specific-heat peaks $C_{L}^{\ast}$, following the scaling law
$C_{L}^{\ast}\sim L^{\alpha/\nu}$. The solid line is a power-law
fit of this form, again for the larger system sizes, and the
corresponding inset illustrates the infinite-volume extrapolation
of the effective exponent ratio $\alpha/\nu$. This analysis leads
to an estimate $\alpha/\nu=0.98(4)$, again very close to the
expected Ising tricritical universality value of
$\alpha/\nu=1$~\cite{lawrie,Blote04}.

\section{Conclusions and outlook}
\label{sec:Conclusions}

In this manuscript, we have presented a numerical study of the
three-dimensional Blume-Capel model defined on a simple cubic lattice.
By implementing a variant of the multicanonical method, we have
performed simulations of the model keeping a constant temperature and
crossing the phase boundary along the crystal-field axis. In this way
we have obtained numerical data for several temperatures in both the
first- and second-order regime of the model, as well as in the
vicinity of the tricritical point. A standard finite-size scaling
analysis, mainly based on a properly defined specific-heat-like
quantity, provided us with precise estimates for the transition points
in both regimes of the phase diagram and with a clear verification of
the expected $\sim L^{d}$ scaling behavior and the Ising universality
class in the first- and second-order regimes of the model,
respectively.

An interesting feature of our study is related to the fact that we
have been able to probe efficiently the low-temperature first-order
regime of the phase diagram of the Blume-Capel model, a rather tricky
numerical task, and obtain accurate estimates of transition points in
the regime of strong crystal fields.  Using the multicanonical method,
and hence simulating otherwise strongly suppressed states, allowed us
to measure the associated free-energy like barrier in the first-order
regime up to the tricritical temperature. This barrier may be related
to the interface tension for spin-$0$ droplets/strips, which we showed
vanishes as one approaches the tricritical point from the first-order
regime.  Moreover, further numerical simulations performed at the
tricritical temperature $T_{\rm t}=1.4182$, proposed by
Deserno~\cite{deserno97}, indicated that this original estimate is
rather accurate, verifying at the same time the expected Ising
tricritical exponent values of $\nu=0.5$ and $\alpha/\nu=1$ from
infinite-volume extrapolations of our effective exponents. 

A further asset of the proposed numerical and scaling schemes is
that it opens a new window for revisiting the effect of disorder
in first-order phase transitions in both two and three dimensions,
where a unified approach to universality is still missing. For
instance, although it is known that in two dimensions under the
presence of bond disorder the ex-second-order regime of the
Blume-Capel model falls into the universality class of the
corresponding random Ising model along the lines of the strong
universality hypothesis~\cite{malakis09}, the same is not true for
the ex-first-order regime. Interestingly enough, for the
ex-first-order regime different results have been obtained for
different lattice geometries~\cite{malakis09,theodorakis12}. The
situation in three-dimensions in even more ambiguous, where one
has to be also careful with respect to the diffused amount of
disorder in the system in order to secure the switching to a
continuous transition~\cite{hui89,berker93}. A recent study of the
random version of the three-dimensional Blume-Capel model
suggested a possible new universality class at the ex-first-order
regime, different to that at the ex-second-order
regime~\cite{malakis12}, an interesting finding if one considers
that the two transitions are between the same ferromagnetic and
paramagnetic phases. Yet, the authors of Ref.~\cite{malakis12}
clearly underlined the need for a more sophisticated approach (in
both numerical and scaling terms) in order to tackle efficiently
the low-temperature disorder-induced continuous transition regime
of the model.

To conclude, using as a platform the Blume-Capel model that shows
the unique feature of having continuous and first-order transition
lines in its phase diagram, we believe that the practise followed
in the present manuscript applied over a wide range of
disorder-strength values and temperatures will provide a better
understanding of the effect of disorder in spin systems. Using the
parallelized version of the multicanonical method and crossing the
phase boundary along the crystal-field axis we expect to be able
to study systematically the universality class and scaling
corrections at the disorder-induced second-order phase transition
of the Blume-Capel model, the shift behavior of the tricritical
point as a function of the disorder strength, and other relevant
open questions. Research in this direction is currently under way.

\begin{acknowledgments}
This project was funded by the European Union and the Free State of Saxony. Part
of this work has been financially supported by the Leipzig Graduate School of
Excellence GSC185 ``BuildMoNa'' and the Deutsch-Franz\"osische Hochschule
DFH-UFA (grant CDFA-02-07).
\end{acknowledgments}

{}


\begin{thebibliography}{}
\bibitem{blume66} M. Blume, Phys. Rev. {\bf 141}, 517 (1966).

\bibitem{capel66} H.W. Capel, Physica (Utr.) {\bf 32}, 966 (1966);
\emph{ibid.} {\bf 33}, 295 (1967); \emph{ibid.} {\bf 37}, 423
(1967).

\bibitem{lawrie} I.D. Lawrie and S. Sarbach, in \emph{Phase Transitions and
  Critical Phenomena}, Vol. 9, edited by C. Domb and J.L. Lebowitz
  (Academic Press, London, 1984).

\bibitem{selke-10} W. Selke and J. Oitmaa, J. Phys. C: Condens. Matter
{\bf 22}, 076004 (2010).

\bibitem{branco} N.S. Branco and B.M. Boechat, Phys. Rev. B {\bf 56}, 11673 (1997).

\bibitem{landau72} D.P. Landau, Phys. Rev. Lett. {\bf 28}, 449
(1972); A.N. Berker and M. Wortis, Phys. Rev. B {\bf 14}, 4946
(1976); M. Kaufman, R.B. Griffiths, J.M. Yeomans and M. Fisher,
\emph{ibid}. {\bf 23}, 3448 (1981); W. Selke and J. Yeomans, J.
Phys. A: Math. and Gen. {\bf 16}, 2789 (1983); D.P. Landau and
R.H. Swendsen, Phys. Rev. B {\bf 33}, 7700 (1986); J.C. Xavier,
F.C. Alcaraz, D. Pena Lara, and J.A. Plascak, \emph{ibid.} {\bf
57}, 11575 (1998).

\bibitem{stephen73} M.J. Stephen and J.L. McColey, Phys. Rev.
Lett. {\bf 44}, 89 (1973); T.S. Chang, G.F. Tuthill, and H.E.
Stanley, Phys. Rev. B {\bf 9}, 4482 (1974); G.F. Tuthill, J.F.
Nicoll, and H.E. Stanley, \emph{ibid.} {\bf 11}, 4579 (1975); F.J.
Wegner, Phys. Lett. {\bf 54A}, 1 (1975).

\bibitem{fox73} P.F. Fox and A.J. Guttmann, J. Phys. C: Condens. Matter {\bf 6},
913 (1973); T.W. Burkhardt and R.H. Swendsen, Phys. Rev. B {\bf
13}, 3071 (1976); W.J. Camp and J.P. Van Dyke, \emph{ibid.} {\bf
11}, 2579 (1975); D.M. Saul, M. Wortis, and D. Stauffer,
\emph{ibid.} {\bf 9}, 4964 (1974).

\bibitem{nightingale82} P. Nightingale, J. Appl. Phys. {\bf 53}, 7927 (1982).

\bibitem{beale86} P.D. Beale, Phys. Rev. B {\bf 33}, 1717 (1986).

\bibitem{jain80} A.K. Jain and D.P. Landau, Phys. Rev. B {\bf 22}, 445 (1980).

\bibitem{landau81} D.P. Landau and R.H. Swendsen, Phys. Rev. Lett. {\bf 46}, 1437 (1981).

\bibitem{care93} C.M. Care, J. Phys. A: Math. and Gen. {\bf 26}, 1481 (1993).

\bibitem{deserno97} M. Deserno, Phys. Rev. E {\bf 56}, 5204 (1997).

\bibitem{Blote95} H.W.J. Bl\"{o}te, E. Luijten, and J.R. Heringa, J. Phys. A: Math. and Gen. {\bf 28}, 6289 (1995).

\bibitem{Blote04} Y. Deng and H.W.J. Bl\"{o}te, Phys. Rev. E {\bf 70}, 046111 (2004).

\bibitem{silva06} C.J. Silva, A.A. Caparica, and J.A. Plascak, Phys. Rev. E {\bf 73}, 036702 (2006).

\bibitem{malakis09} A. Malakis, A.N. Berker, I.A. Hadjiagapiou, and N.G. Fytas, Phys. Rev. E {\bf 79}, 011125
(2009); A. Malakis, A.N. Berker, I.A. Hadjiagapiou, N.G. Fytas,
and T. Papakonstantinou, \emph{ibid.} {\bf 81}, 041113 (2010).

\bibitem{malakis12} A. Malakis, A.N. Berker, N.G. Fytas, and T.
Papakonstantinou, Phys. Rev. E {\bf 85}, 061106 (2012).

\bibitem{theodorakis12} P.E. Theodorakis and N.G. Fytas, Phys. Rev. E {\bf 86}, 011140 (2012).

\bibitem{fytas} N.G. Fytas, Eur. Phys. J B {\bf 79}, 21 (2011).

\bibitem{fytaspanos} N.G. Fytas and P.E. Theodorakis, Eur. Phys. J. B {\bf 86}, 30 (2013).

\bibitem{berg1992} B.A. Berg and T. Neuhaus, Phys. Lett. B {\bf 267}, 249 (1991); Phys. Rev. Lett. {\bf 68}, 9 (1992).

\bibitem{janke1998} W. Janke, Int. J. Mod. Phys. C {\bf 03}, 1137 (1992); Physica A {\bf 254}, 164 (1998).

\bibitem{zierenberg2013} J. Zierenberg, M. Marenz, and W. Janke, Comput. Phys. Comm.
{\bf 184}, 1155 (2013); Physics Procedia {\bf 53}, 55 (2014).

\bibitem{reweighting} W. Janke, {\it Monte Carlo methods in classical statistical physics},
invited lectures, in: {\it Computational Many-Particle Physics},
edited by H. Fehske, R. Schneider, and A. Wei{\ss}e, Lect. Notes
Phys. 739 (Springer, Berlin, 2008); pp. 79-140.

\bibitem{WL} F. Wang and D.P. Landau, Phys. Rev. Lett. {\bf 86}, 2050 (2001); Phys. Rev. E {\bf 64}, 056101 (2001).

\bibitem{troyer03} M. Troyer, S. Wessel, and F. Alet, Phys. Rev. Lett. {\bf 90},
120201 (2003).

\bibitem{dayal04} P. Dayal, S. Trebst, S. Wessel, D. W\"{u}rtz, M. Troyer, S. Sabhapandit, and S.N.
Coppersmith, Phys. Rev. Lett. {\bf 92}, 097201 (2004).

\bibitem{belardinelli07} R.E. Belardinelli and V.D. Pereyra, Phys. Rev. E {\bf 75}, 046701 (2007).

\bibitem{fytasme} N.G. Fytas, A. Malakis, and K. Eftaxias, J.
Stat. Mech.: Theory Exp. (2008) P03015.

\bibitem{Schoebl2011} S. Sch{\"o}bl, J. Zierenberg, and W. Janke, Phys. Rev. E {\bf 84}, 051805 (2011).

\bibitem{hartmann} A.K. Hartmann and A.P. Young, Phys. Rev. B {\bf 64}, 180404
(2001).

\bibitem{efron1982} B. Efron, {\it The Jackknife, the Bootstrap and other Resampling Plans}
(Society for Industrial and Applied Mathematics, Philadelphia,
1982).

\bibitem{hasen} M. Hasenbusch, Phys. Rev. B {\bf 82}, 174434 (2010).

\bibitem{BergMulti}
  B.~A. Berg and W. Janke, Phys. Rev. Lett. {\bf 98}, 040602 (2007);
                           Physics Procedia {\bf 7}, 19 (2010).
\bibitem{Bittner2011}
  E. Bittner and W. Janke, Phys. Rev. E {\bf 84}, 036701 (2011).

\bibitem{barber83} M.N. Barber, in \emph{Phase Transitions and Critical Phenomena}, Vol. 8, edited by C. Domb and J.L. Lebowitz
(Academic, NY, 1983).

\bibitem{Nussbaumer2006} A. Nu\ss baumer, E. Bittner, T. Neuhaus, and W. Janke, Europhys.
Lett. {\bf 75}, 716 (2006).

\bibitem{Nussbaumer2008} A. Nu\ss baumer, E. Bittner, and W. Janke,  Phys. Rev. E {\bf 77}, 041109 (2008).

\bibitem{Bittner2009} E. Bittner, A. Nu\ss baumer, and W. Janke, Nucl. Phys. B {\bf
820}, 694 (2009).

\bibitem{guida} R. Guida and J. Zinn-Justin, J. Phys. A: Math. and Gen. {\bf 31}, 8103 (1998).

\bibitem{hui89} K. Hui and A.N. Berker, Phys. Rev. Lett. {\bf 62}, 2507 (1989); \emph{ibid.} {\bf 63}, 2433(E) (1989).

\bibitem{berker93} A.N. Berker, Physica A {\bf 194}, 72 (1993).

\end{thebibliography}
\end{document}